\documentclass[acmsmall,screen,nonacm]{acmart}
\AtBeginDocument{%
  }

\setcopyright{acmlicensed}
\copyrightyear{2018}
\acmYear{2018}
\acmDOI{XXXXXXX.XXXXXXX}
\acmConference[Conference acronym 'XX]{Make sure to enter the correct
  conference title from your rights confirmation email}{June 03--05,
  2018}{Woodstock, NY}
\acmISBN{978-1-4503-XXXX-X/2018/06}

\usepackage{acronym}
\usepackage{comment}
\usepackage{todonotes}
\usepackage{xspace}
\usepackage{csquotes}
\usepackage{semantic}
\usepackage[inline]{enumitem}
\usepackage[nameinlink,capitalise]{cleveref}
\crefname{example}{Example}{Examples}
\Crefname{example}{Example}{Examples}
\usepackage[normalem]{ulem}
\usetikzlibrary{shapes,arrows,positioning,backgrounds}
\usepackage{subcaption}
\usepackage{booktabs}
\usepackage{multirow}
\usepackage{array}

\begin{document}

\newacro{TRS}{Term Rewriting System}
\newacro{KBO}{Knuth-Bendix Order}
\newacro{KBC}{Knuth-Bendix Completion}
\newacro{EqSat}{Equality Saturation}
\newacro{CP}{Critical Pair}
\newacro{AST}{Abstract Syntax Tree}

\title{Augmenting Rewrite Rule Sets via Knuth-Bendix Completion}

\author{Michael Schifferer}
\email{misc00021@stud.uni-saarland.de}
\affiliation{
    \institution{Saarland University, Saarland Informatics Campus}
    \city{Saarbrücken}
    \country{Germany}
}

\author{Marcel Ullrich}
\email{ullrich@cs.uni-saarland.de}
\orcid{0009-0006-0127-9623}
\affiliation{
    \institution{Saarland University, Saarland Informatics Campus}
    \city{Saarbrücken}
    \country{Germany}
}

\author{Sebastian Hack}
\email{hack@cs.uni-saarland.de}
\orcid{0000-0002-3387-2134}
\affiliation{
    \institution{Saarland University, Saarland Informatics Campus}
    \city{Saarbrücken}
    \country{Germany}
}

\begin{abstract}
\ac{EqSat} is a powerful technique for program optimization, systematically exploring the search space of candidate programs to overcome the phase ordering problem. 

However, the feasibility and performance of EqSat rely heavily on the specific rewrite rules used to derive equivalent programs. 
These rule sets are typically handcrafted, requiring extensive domain expertise and carrying the risk of missing valuable transformations. 

In this work, we evaluate \ac{KBC} as a method for automatically generating and augmenting rewrite rules for EqSat.
Our experiments show faster optimization as well as reaching better terms with previously missed optimizations.

\end{abstract}




\received{20 February 2007}
\received[revised]{12 March 2009}
\received[accepted]{5 June 2009}

\def\eg{e.g.\xspace}
\def\ie{i.e.\xspace}

\maketitle

\section{Introduction}
\label{sec:intro}

Optimizing compilers rely heavily on program transformations to improve code performance. However, applying these transformations sequentially introduces the phase ordering problem, where the order of applied optimizations heavily limits the final program quality. \ac{EqSat} addresses this challenge by representing multiple equivalent programs simultaneously within a specialized data structure known as an e-graph. By systematically applying transformations without committing to a single path, \ac{EqSat} exhaustively explores the search space of candidate programs to find the optimal representation.

While \ac{EqSat} proves to be an effective optimization framework, its feasibility and performance depend on the underlying rewrite rules used to derive equivalent programs. Compiler developers handcraft these rule sets to carefully balance the trade-off between enabling a deep search space and preventing the e-graph from growing out of control. Designing these rule sets requires strong domain knowledge, intuition, and extensive testing. Because manual rule creation is prone to human error, developers often miss valuable, non-obvious transformations. Furthermore, little is known about the exact properties of highly effective rule sets, and the automated tools available to generate them remain limited.

To address this limitation, we evaluate \acf{KBC}, a well-established procedure from automated theorem proving, as a method for automatically generating and augmenting rewrite rules for program optimization. We develop a methodology to process an initial, handwritten rule set using \ac{KBC} and evaluate the resulting rules in two distinct environments. First, we test the \ac{KBC}-generated rules within the standard \ac{EqSat} framework to see if they can uncover previously missed optimizations. Second, we evaluate them using a custom greedy rewriting engine, which trades the exhaustive exploration of \ac{EqSat} for compile time and memory efficiency at the risk of result programs of inferior quality.

Our experiments show that augmenting handwritten rules with KBC-generated rules results in faster optimization and allows the system to reach better terms with previously missed optimizations.
In this paper, we present
\begin{itemize}
    \item different methods for generating and processing rewrite rule sets for program optimization using \ac{KBC},
    \item an experimental evaluation demonstrating that \ac{KBC}-augmented rule sets improve equality saturation in terms of overall simplification effectiveness and running time compared to strictly handwritten rules, and
    \item analyze the viability of greedy rewriting as a resource-efficient alternative to \ac{EqSat}, finding that while it minimizes memory and time usage, it still achieves considerably less effective simplification than \ac{EqSat}.
\end{itemize}

\Cref{sec:background} establishes the background on term rewriting, equality saturation, and Knuth-Bendix completion. \Cref{sec:main} details our methodology for rule generation and the experimental setup. \Cref{sec:eval} presents our results across both the EqSat and greedy rewriting frameworks. Finally, we discuss related work in \Cref{sec:relwork} and our conclusion in \Cref{sec:conclusion}.

\section{Term Rewriting Systems}
\label{sec:background}

Given a \emph{simplification order} $\succ$ on terms, \ac{KBC} turns a set of equalities into a \emph{confluent} and \emph{terminating} \ac{TRS}. If successful, it ensures that equivalent terms ($t=_E^\ast t'$) simplify to a unique normal form.

A simplification order $\succ$ requires three properties:
\begin{enumerate}
    \item Compatibility: $s\succ t \implies f(\dots,s,\dots)\succ f(\dots,t,\dots)$ for all $f\in\Sigma$.
    \item Stability: $s\succ t \implies s\sigma\succ t\sigma$ for any substitution $\sigma$.
    \item Subterm ordering: If $s$ is a proper subterm of $t$, then $t\succ s$.
\end{enumerate}

A common implementation is the \ac{KBO}, which weighs terms based on predefined symbol weights, breaking ties using operator precedence and lexicographic checks.

\ac{KBC} is driven by two main concepts:
First, \emph{orientation}: an equality $s=t$ permanently becomes a directed rewrite rule $s\to t$ if $s\succ t$. 
Second, \emph{critical pairs}: if a term $u$ diverges into two different terms ($s\leftarrow_R u\to_R t$), we deduce $s=t$. This creates a new rule to bridge them. Visually, this creates a shortcut that \enquote{tunnels through peaks} in a proof, ensuring confluence.

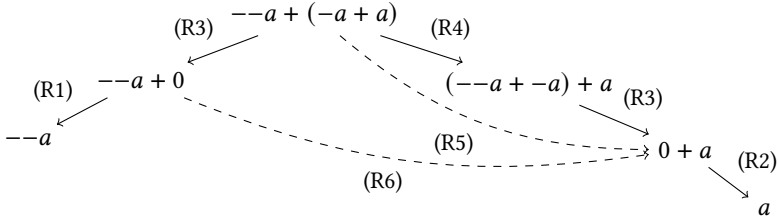
\begin{figure}[htpb]
\begin{center}
\begin{tikzpicture}[node distance=5mm]
   \node (n2) at (0, 0) {$-{-a}+(-a+a)$} ;
   \node (n1) [below left=of n2] {$-{-a}+0$} ;
   \node (n0) [below left=of n1] {$-{-a}$} ;
   \node (n3) [below right=of n2] {$(-{-a}+-a)+a$} ;
   \node (n4) [below right=of n3] {$0+a$} ;
   \node (n5) [below right=of n4] {$a$} ;
   \draw[->] (n2) -- node[midway, left,  anchor=south east, font=\small] {(R3)} (n1);
   \draw[->] (n1) -- node[midway, left,  anchor=south east, font=\small] {(R1)} (n0);
   \draw[->] (n2) -- node[midway, right, anchor=south west, font=\small] {(R4)} (n3);
   \draw[->] (n3) -- node[midway, right, anchor=south west, font=\small] {(R3)} (n4);
   \draw[->] (n4) -- node[midway, right, anchor=south west, font=\small] {(R2)} (n5);
   \draw[->] (n2) edge[dashed,bend right=20] node[midway, left, anchor=north east, font=\small] {(R5)} (n4);
   \draw[->] (n1) edge[dashed,bend right=15] node[midway, left, anchor=north east, font=\small] {(R6)} (n4);
\end{tikzpicture}
\end{center}
\caption{A rewrite proof where critical pairs tunnel the peaks.}
\label{fig:kbc:proof}
\end{figure}

\ac{KBC} continuously applies elementary inference rules to a set of equalities~$E$ and rewrite rules~$R$:
\[
\inference[Orient]{
s\succ t \\
E\cup\{s=t\},\ R 
}{
E,\ R\cup\{s\to t\}
}
\qquad
\inference[Delete]{
E\cup\{s=s\},\ R 
}{
E,\ R
}
\qquad
\inference[Deduce]{
s\leftarrow_R u\to_R t \\
E,\ R 
}{
E\cup\{s=t\},\ R
}
\]

A standard \ac{KBC} run results in success (yielding a confluent \ac{TRS}), failure (due to unorientable equations like $X+Y=Y+X$), or non-termination.

\section{Approach}
\label{sec:main}

To automate the generation of rewrite rules for program optimization, we design a pipeline consisting of two primary phases: synthesizing rule sets via \ac{KBC} and deploying these rule sets across two distinct rewriting environments.

\textbf{Rule Generation.} 
We use the twee \citep{twee} theorem prover to perform \ac{KBC} on an initial set of handwritten arithmetic axioms.
Twee derives \acp{CP}, augmenting the original rule set.

For unorderable rules, we investigate removing them vs keeping both orientations, left to right and right to left, as two rules.
We create multiple rule set variants by using different completion limits (the maximum number of rules generated) and by evaluating whether to replace the original handwritten rules entirely or augment them with the newly generated rules. 
Finally, we isolate and process partial conditional operations (such as division) separately to ensure that the generated rules remain sound and avoid invalid states like division by zero.

The rules produced by \ac{KBC} are oriented. However, keeping the original rules is often beneficial as \ac{EqSat} uses expanding rules to explore a large state space of terms for possible optimization.
We categorize the \ac{KBC} rules into three categories: 
\begin{itemize}
    \item Normal rules: Rules that are already present in the original rule set or specializations of present rules
    \item Shortcut rules: Rules that result from combining multiple rule applications into a single step
    \item Novel rules: Rewrites that are not achieved by \ac{EqSat} as they require inventing a value. Such a rule is $y\cdot (y^{-1}\cdot z) \to 1\cdot z$ from the rules $x\cdot x^{-1}\to 1$ and $(x\cdot y)\cdot z \to x\cdot (y\cdot z)$. To reach this rewrite, a value for $x$ in $1=x\cdot x^{-1}$ has to be invented.
\end{itemize}

Special care in completion of rules is needed for partial functions with preconditions like division.
The rules $0/x=0$ and $x/x=1$ allow proving $1=0$ for $x=0$.
Therefore, conditions need to be added that restrict instantiation with $0$.
We use the encoding defined by \citet{ClaessenSmallbone2021} to encode preconditions as \texttt{if-then-else} constructs.
The egg rule \texttt{rw!("cancel-div"; "(/ ?x ?x)" => "1" if is\_not\_zero("?x"))} is expressed as \texttt{ifeq(not\_zero(x),true,x/x,1)$\to$ 1}.

\textbf{Rewriting Environments.} 
Once we generate and process the rule sets, we deploy them into two distinct optimization frameworks to evaluate their performance. 

First, we integrate the rules into a standard \ac{EqSat} engine. This environment leverages the e-graph data structure to exhaustively explore equivalent program representations simultaneously. By applying all possible rules at every step without committing to a single rewrite path, this framework allows us to observe the maximum simplification power of our \ac{KBC}-generated rules.

Second, we apply the exact same rule sets within a custom greedy rewriting engine. Unlike \ac{EqSat}, the greedy approach does not explore multiple paths. Instead, it applies simplifying rules sequentially and deterministically until no further rule matches exist.
We select rules according to the generation order defined by Twee. Twee enumerates rules heuristically, starting with rules with few operators.
By deploying the rules in both frameworks, we isolate the impact of the rule sets themselves and directly contrast the exhaustive, resource-intensive nature of \ac{EqSat} against the lightweight, sequential execution of greedy rewriting.

\ac{KBC} produces a confluent rewrite system if it terminates. Such a rewrite system is desired as rules can be applied in an arbitrary order.
We hypothesize that a partially completed rewrite system by \ac{KBC} exhibits confluent properties allowing to apply it greedily.


\section{Experimental Evaluation}
\label{sec:eval}

We evaluate our \ac{KBC}-generated rule sets based on mean symbol reduction, the average decrease in the total number of operators between the input and the optimized terms. We conduct our experiments across both the exhaustive \ac{EqSat} engine and our custom greedy rewriting engine, testing randomly generated arithmetic terms of varying sizes (small=$1000$ terms with $3$ binary operators, large=$500$ terms with $10$ binary operators, and huge=$250$ terms with $25$ binary operators). 
We use the arithmetic rewrite rules from \citet{egg_math_rules}.

\textbf{Equality Saturation Performance.} 
First, we evaluate how \ac{KBC}-generated rules affect the simplification power of \ac{EqSat}. We observe that strictly replacing the handwritten rules with \ac{KBC}-generated rules often causes the system to get stuck in local optima, because \ac{KBC} naturally filters out the complexity-increasing rules that \ac{EqSat} relies on to bridge equivalent states. For instance, $(x^3-1)/(x-1)$ first needs to be expanded to $(x^3-x^2+x^2-x+x-1)/(x-1)$ before it can be simplified to $x^2+x+1$.
The original rules contain rules in both directions to expand terms and reduce them.
\ac{KBC} only keeps the reducing rules in its rule set.

\begin{figure}
\begin{subfigure}{.49\textwidth}
    \centering
    \includegraphics[width=\linewidth]{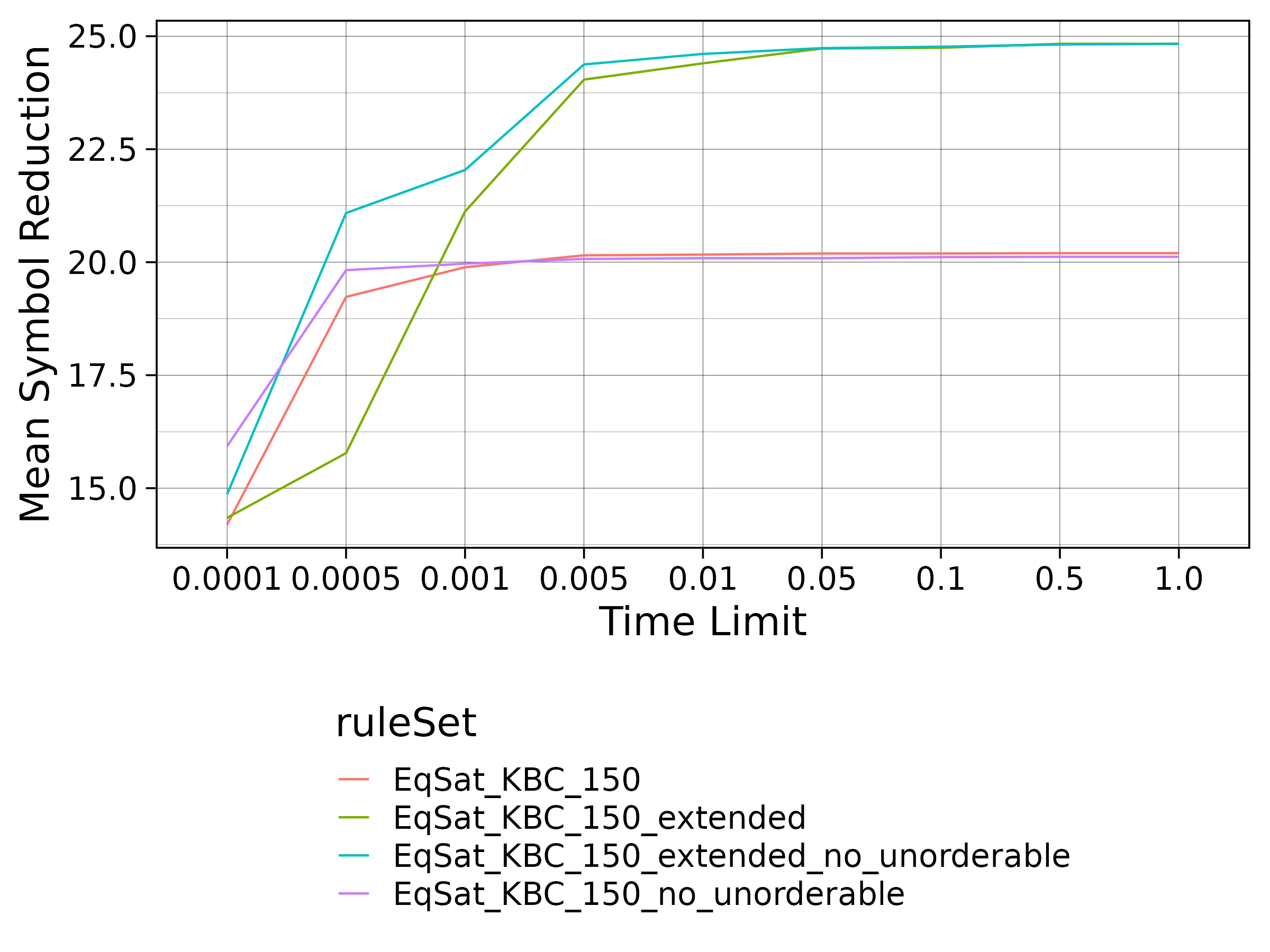}
    \caption{Impact of rule set postprocessing on mean symbol reduction for huge terms.}
    \label{fig:postprocessing_huge}
\end{subfigure}\hspace{.01\textwidth}
\begin{subfigure}{.49\textwidth}
    \centering
    \includegraphics[width=\linewidth]{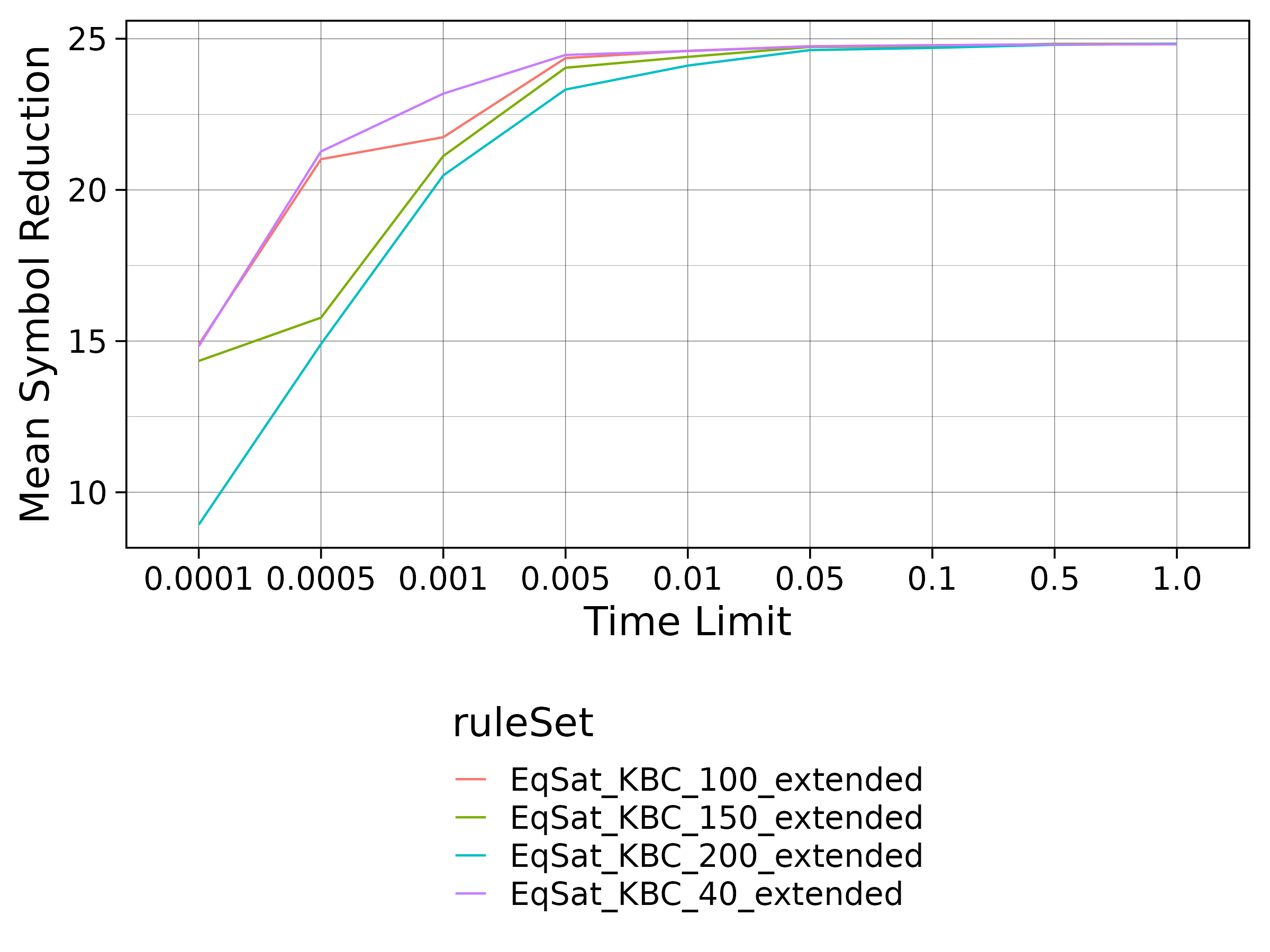}
    \caption{Simplification effectiveness across different time limits relative to the number of generated rules used to extend the base set.}
    \label{fig:rule_number_extended}
\end{subfigure}%
\end{figure}

On the other hand, extending the original handwritten rules with \ac{KBC}-generated rules yields consistently better results as seen in \Cref{fig:postprocessing_huge}.

For example, on large terms containing partial functions, the \ac{KBC}-augmented rule set (\texttt{E-KBC150}) achieves a mean symbol reduction of $10.46$, outperforming the strictly handwritten baseline (\texttt{E-base}) at $9.73$, see \Cref{tab:framework_comparison}. On terms without partial functions, \texttt{E-KBC150} reaches a reduction of $14.13$ compared to the baseline's $13.61$. We find that generating up to $150$ additional rules maximizes effectiveness; beyond this threshold, larger rule sets slow down the \ac{EqSat} convergence without providing meaningful improvements in term reduction as shown in \Cref{fig:rule_number_extended}. 


\textbf{Greedy Rewriting Performance.} 
We evaluate the rule sets within a greedy rewriting environment, where the engine applies rules sequentially without tracking multiple term variations. In this setting, the \ac{KBC}-generated rule sets vastly outperform the original handwritten rules. The handwritten rules rely heavily on \ac{EqSat}'s ability to increase term size for exploration. Therefore, the handwritten rules fail to simplify terms in a greedy, sequential setup. In contrast, the \ac{KBC} rules provide directly simplifying paths. Unlike in \ac{EqSat}, greedy rewriting benefits significantly from much larger rule sets; we observe that simplification effectiveness continuously improves as we increase the rule set size up to 1,000 rules (\texttt{G-KBC1000}), after which the benefits begin to plateau. 

\textbf{Cross-Framework Comparison.} 
Finally, we compare the best-performing \ac{KBC} configurations of both frameworks to highlight the trade-off between optimization power and resource efficiency.  \ac{EqSat} consistently achieves higher overall simplification effectiveness, outperforming the greedy approach in mean symbol reduction. 

\begin{table}[htpb]
    \centering
    \caption{Comparison of optimal rule sets across EqSat and greedy rewriting frameworks on small terms.}
    \label{tab:framework_comparison}
    \begin{tabular}{llrr}
        \toprule
        \textbf{Term Set} & \textbf{Configuration} & \textbf{Mean Reduction} & \textbf{Time (s)} \\
        \midrule
        \multirow{2}{10em}{without div/pow, large}
            & G-KBC1000 & 10.88  & 0.00009 \\
            & E-KBC100  & \textbf{14.13} & 1.0000 \\
            & E-base    & 13.61  & 0.5 \\
        \midrule
        \multirow{3}{10em}{with div/pow, large}
            & G-KBC1000 & 8.92  & 0.000081 \\
            & E-KBC150  & \textbf{10.46} & 0.05 \\
            & E-base    & 9.73  & 0.005 \\
        \bottomrule
    \end{tabular}
\end{table}

However, greedy rewriting requires several orders of magnitude less time to reach its optimal solution. For instance, as shown in \Cref{tab:framework_comparison}, on large terms with partial functions, \texttt{G-KBC1000} executes in just 0.000081 seconds, whereas \texttt{E-KBC150} requires 0.05 seconds. Ultimately, our experiments confirm that \ac{KBC} successfully enhances \ac{EqSat} for maximum optimization power, while simultaneously transforming greedy rewriting into a efficient alternative with significantly less optimization time and memory consumption.

\section{Related Work}
\label{sec:relwork}

To accurately position our contributions, we discuss notable work in equality saturation, automated rewrite rule generation, and specialized e-graph variations.

\textbf{Equality Saturation.} \ac{EqSat} was initially introduced to solve the phase ordering problem by systematically exploring equivalent programs~\cite{Tate_2011}. The approach gained significant traction with the introduction of the \texttt{egg} framework~\citep{Willsey_2021}, which drastically improves extensibility via domain-specific analyses and performance through e-graph rebuilding. Today, researchers apply \ac{EqSat} to diverse domains, including floating-point error reduction in \texttt{Herbie}~\citep{herbie} and tensor program optimization in \texttt{Felix}~\citep{Felix}. However, these practical applications rely almost exclusively on manually constructed rewrite rules, highlighting the need for the automated synthesis techniques explored in this work.

\textbf{Rewrite Rule Synthesis.} Recent advancements aim to ease the burden of manual rule creation and improve \ac{EqSat} workflows (e.g., fast relational e-matching~\citep{Willsey_2022} and \texttt{SmoothE} for complex term extraction~\citep{SmoothE}). For rule generation specifically, \texttt{Ruler} infers rewrite rules by enumerating terms within an e-graph and merging equivalence classes based on empirical behavior across input vectors~\citep{Ruler}. While \texttt{Ruler} successfully synthesizes minimal rule sets, its rules only guarantee empirical equivalence within \ac{EqSat} and lack the formal confluence properties provided by completion algorithms. A recent survey on rewrite rule synthesis acknowledges \ac{KBC}~\citep{10.1093/comjnl/34.1.2} as a foundational technique but notes the lack of evaluation regarding its practical effectiveness in program optimization~\citep{Hong_2024}. Our work directly addresses this gap.

\textbf{Acyclic E-Graphs.} Recent variations of the e-graph data structure, such as \ae graphs used in the Cranelift compiler, maintain strict acyclicity to enable fast, bidirectional conversions with control flow graphs~\citep{Fallin2023_aEGraphs}. Because \ae graphs disallow cyclic derivations, they enforce a strict directionality requirement for most rewrite rules. This constraint aligns perfectly with \ac{KBC}'s focus on strictly directed rewrites, suggesting that \ac{KBC}-generated rule sets may be uniquely well-suited for optimization passes within acyclic e-graph frameworks.

\section{Conclusions}
\label{sec:conclusion}

In this work, we evaluate \ac{KBC} as a systematic method for automatically augmenting rewrite rules for program optimization. 
Our experimental results demonstrate that augmenting handwritten rule sets with \ac{KBC}-generated rules significantly improves the simplification speed and effectiveness of \ac{EqSat}. 

The critical pairs generated by \ac{KBC} act as a shortcut of rewrite chains, as well as providing novel rewrites previously not expressed by the directed nature of \ac{EqSat} rule applications.
However, adding too many critical pair rules can significantly increase the egraph size, thereby decreasing performance.

Furthermore, our evaluation reveals a clear trade-off with greedy rewriting: while it executes several orders of magnitude faster than \ac{EqSat}, it achieves significantly lower simplification effectiveness under the tested conditions. 
Ultimately, \ac{KBC} proves to be a valuable tool for enhancing \ac{EqSat} and reducing the manual burden of rule engineering.

\subsection{Future Work}
\label{sec:future-work}

This study establishes a foundation for automated rewrite rule synthesis using completion techniques, opening several avenues for future research.
\begin{itemize}
    \item Evaluate \ac{KBC} on other domains such as bitvectors
    \item Use \ac{KBC} during EGraph saturation to find useful critical pairs
    \item Heuristics on helpful critical pairs generated by \ac{KBC}
    \item Automatic rule generation with properties that make \ac{KBC} superfluous
\end{itemize}




\bibliographystyle{ACM-Reference-Format}
\bibliography{references}




\end{document}